\documentclass[english]{llncs}
\usepackage[T1]{fontenc}
\usepackage[latin9]{inputenc}
\usepackage{units}
\usepackage{graphicx}

\makeatletter
\@ifundefined{showcaptionsetup}{}{%
 \PassOptionsToPackage{caption=false}{subfig}}
\usepackage{subfig}
\makeatother

\usepackage{babel}

\begin{document}

\title{A Parallelism-Based Approach to Network Anonymization}

\author{Igor Margasi\'{n}ski}

\institute{Institute of Telecommunications, Warsaw University of Technology\\
\texttt{igor@tele.pw.edu.pl}}
\maketitle
\begin{abstract}
Considering topologies of anonymous networks we used to organizing
anonymous communication into hard to trace paths hiding its origin
or destination. In anonymity the company is crucial, however the serial
transportation imposes a costly tradeoff between a level of privacy
and a speed of communication. 

This paper introduces a framework of a novel architecture for anonymous
networks that hides initiators of communications by parallelization
of anonymous links. The new approach, which is based on the grounds
of the anonymous P2P network called P2Priv, does not require content
forwarding via a chain of proxy nodes to assure high degree of anonymity.
Contrary to P2Priv, the new architecture can be suited to anonymization
of various network communications, including anonymous access to distributed
as well as client-server services. In particular, it can be considered
as an anonymization platform for these network applications where
both privacy and low delays are required.
\end{abstract}

\section{Introduction}

Anonymous communications by means of public packet networks involve
two contradictory tasks: the first is \textbf{transport and routing}---which
requires a detail information on origin and destination points of
communications, and the second is \textbf{anonymization}---which is
basically aimed at hiding of this information and especially an association
between them. Certainly, addressing information is essential for a
successful delivery of content and therefore it cannot be expunged.
Hence in general, the ally of anonymous communication is collective
\cite{usability:weis2006,danezis:weis2006,mixmaster-reliable}. The
single word \emph{crowd} speaks volumes about anonymity. The more
numerous a set of actors involved in an information exchange\emph{
blending into a crowd} is easier to achieve. And then, the higher
traffic volume among these actors the faster our traffic can be hidden
and exchanged. In general, to represent such a collective, an operation
of anonymous networks is based on a sequential traffic forwarding
by a subgroup of network nodes, also known as \emph{proxy chaining}
\cite{onion-routing:ih96}. However, substantial delays mount up in
this way. This paper introduces an alternate architecture for anonymous
networks, a network privacy preserving parallel topology, where network
actors organize themselves in parallel links. The topology of the
new solution evolves from the anonymous P2P network called P2Priv
\cite{MarPi08}. However, the applications of the new architecture
are not limited to P2P content distribution and its deployment can
be considered for generic-purpose anonymous networks. 

The rest of the paper is organized as follows. Section 2 provides
an introduction to topology issues of anonymous networks. Section
3 describes a model of an adversary while Section 4 contains an anonymity
analysis of parallel architecture of P2Priv applied first, in accordance
with its intended use, for P2P content distribution, and secondly,
for client-server like scenarios. Section 5 contains a description
of our novel solution able to assure anonymity for centralized network
services, while its anonymity analysis is presented in Section 6.
Conclusions and discussion of a future work are included in Section
7.

\section{Background}

The topology of anonymous networks has been widely discussed since
the introduction of Chaum's Mix-net anonymous network \cite{chaum-mix}
and among a variety of anonymous networks, the most attention has
been devoted to the interconnection issues of Mix-based nodes. Originally,
the route through a cascade of Mixes was fixed. Further improvements
allowed a sender to randomly select a path for each message in the
so called free-route topologies \cite{disad-free-routes,sync-batching}.
Hybrid models with a restricted number of connections and path selection
narrowed to restricted-routes were proposed in \cite{danezis:pet2003}.
Both fixed cascades and fully interconnected Mix networks with random
routes have assorted constraints and the advantage of each depends
primarily on the area of their deployment and the scale of the network
\cite{disad-free-routes,mixcascades,cottrell,babel}. Today's Mixes
allow to route content in various ways determined by sender nodes. 

Besides Mixes, other designs of anonymous networks were proposed within
individual interconnection strategies. Anonymous message-by-message
routing called \emph{onion routing} was proposed by Goldschlag et
al. \cite{onion-routing:ih96}. Today, several implementations of
this concept are available, including the low-latency network called
\emph{the second generation onion router} (TOR) designed by Dingledine
et al. \cite{dingledine:tor}---the general purpose anonymous network
of the world-wide range.

Reiter et al. proposed other low-latency anonymous network called
\emph{Crowds }with anonymous routing based on the\emph{ rando-walk}
step \cite{crowds:tissec}. In this strategy senders do not influence
a path selection which is determined in a random manner in each hop
sequentially. Both hop-by-hop and message-by-message routing strategies
are less robust against traffic dropping by proxy nodes than Mix-network.
Still, their simplicity makes them attractive in practice \cite{camlys05}. 

By and large, the common feature of anonymous networks is their \textbf{serial
}architecture and a formation of untraceable paths via middle-man
nodes for anonymous content forwarding. Due to such praxis, and in
combination with traffic encryption and anonymization techniques deployed
inside proxy nodes (e.g., content batching, aggregations, and permutations),
an observer of the anonymous networks cannot practically point out
senders and receivers in a batch of intermixed content flows via middle-man
nodes.

\subsection{Parallel Architecture of P2Priv}

\begin{figure}
\begin{centering}
\includegraphics[scale=0.44]{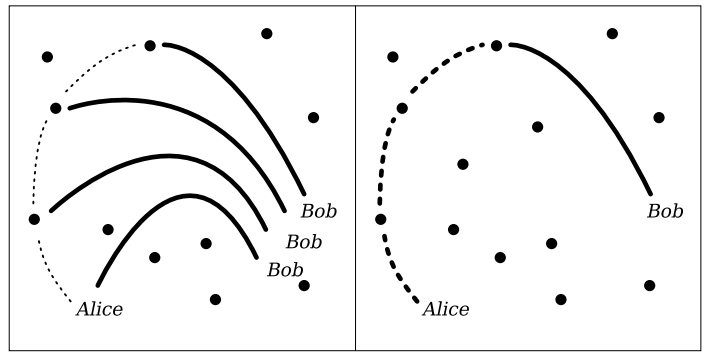}
\par\end{centering}

\caption{\label{fig:P2PRIV-architecture}P2Priv architecture (\emph{left})
as compared to classical anonymous network (\emph{right}). }

\end{figure}

The serial content forwarding, known from today's anonymous networks,
was omitted in the architecture of the anonymous peer-to-peer overlay
network called P2Priv (peer-to-peer direct and anonymous distribution
overlay) proposed by Margasi\'{n}ski et al. in \cite{MarPi08}. The
topology of the P2Priv network, in respect to content transportation,
involves a number of additional virtual links similar to classical
anonymous networks; however it is arranged in a \textbf{parallel}
manner. Figure \ref{fig:P2PRIV-architecture} illustrates the parallel
architecture of P2Priv with solid lines representing plain-text communication
and dotted lines corresponding to communication secured by the anonymization
techniques. Let us have a closer look at the P2Priv architecture.
Before a content transportation, a signalization token with meta-data
describing the content is forwarded over classical anonymous paths
towards formation of so called \emph{cloning cascades} ($CC$). The
well-known anonymous techniques (i.e., Mix network and random walk
algorithm) are utilized in the anonymization process of this \emph{lightweight}
communication (traffic comprised by numerous and short messages sent
by random nodes is in favour of a Mix-net performance) and hiding
the initiator of the $CC$. Then, after a random interval of time,
each $CC$ member (i.e., group of the so called \emph{clones} and
the true initiator) communicates directly and independently with a
destination node or nodes towards content transportation. The anonymity
of P2Priv is based on a collective formed in a parallelism-based manner,
as every content exchange in P2Priv is accompanied by a simultaneously
generated exchange of the same content between random nodes. The process
of finding the true initiator among P2Priv nodes is hard to perform
even for an adversary able to collude a significant range of nodes.

Results of anonymity and traffic performance analysis are promising
for P2Priv \cite{MarPi08,P2PRIVtrans}. However, they have been obtained
for a distributed environment which is not always available in general-purpose
communications.

\section{Threat Model}

We consider a threat model with a partial adversary who controls a
colluding fraction $\rho$ of all overlay network nodes $N$. These
malicious nodes are able to provide both passive and active attacks.
We consider a static adversary unable to arbitrarily adapt the set
of malicious nodes among user nodes and we assume that colluding nodes
are uniformly distributed among equivalent nodes. However, whenever
we will deal with server-like nodes or just centralized points of
communication, we will treat these nodes as easy to choose points
of observation and assume that these nodes can be compromised. Using
information entropy measures and based on the information theoretic
anonymity measurement model \cite{Diaz02,Serj02}, an adversary who
posses no information about the system can describe his/her uncertainty
in successful finding of the initiator of a particular action (let
as call her \emph{Alice}) as \begin{eqnarray}
\mathcal{H}_{\max} & = & -\sum_{i=1}^{\left|N\right|(1-C_{s})}\frac{1}{\left|N\right|(1-\rho)}\log_{2}(\frac{1}{\left|N\right|(1-\rho)}).\label{eq:}\\
 & = & -\log_{2}(\left|N\right|(1-\rho)).\nonumber \end{eqnarray}

\section{Anonymity Analysis for P2Priv}

As the new architecture proposed in this paper share some primitives
with P2Priv, we will apply the threat model to P2Priv architecture
first and discuss its anonymity in two service scenarios. In the first
one, likely to occur in fully distributed P2P systems, we will consider
content distribution (e.g., P2P file-sharing) among equivalent peers.
Secondly, we will analyze anonymity offered by P2Priv for typical
client-server network services where client nodes connect repeatedly
to some server (e.g., a popular Web server). Notice that models, similar
to the first scenario, has been investigated previously in \cite{MarPi08,IM-thesis,P2PRIVtrans}
with different adversary possibilities considered. The second scenario
has not been considered yet. For simplicity purposes, the scenarios
will be referred to as \textbf{P2P Scenario} and \textbf{Client-Server
Scenario} respectively.

The detailed description of P2Priv was introduced in\cite{P2PRIVtrans}.
At a glance, cloning cascade ($CC$) of P2Priv is established in random-walk
manner with a mean length described by probability $p_{f}$, as follows
\cite{P2PRIVtrans}

\begin{center}
\begin{equation}
\left|CC\right|=\frac{p_{f}-2}{p_{f}-1}.\label{eq:-8}\end{equation}

\par\end{center}

Each step of the token's random-walk is sent by a Mix-network layer
formed for this purpose by user nodes ($N$). After establishing $CC$,
randomly delayed direct connections originate from $CC$ nodes towards
content transportation (Figure \ref{fig:P2PRIV-architecture}). The
analysis of various attacks on P2Priv (presented in \cite{IM-thesis})
shows that a secure configuration of P2Priv starts from $p_{f}=2/3$
which corresponds to $CC$ mean length equal 4.

\subsection{P2Priv in P2P Scenario}

The adversary tries to find out who initiates a content distribution
in respect to some content of his interest. A linkage between the
cloning token's sender---\emph{Alice} and the cloning token is hidden
by means of the Mix-net layer. Mix-net is recognized as one of the
strongest anonymization methods. However, it requires adequate user
traffic characteristics to achieve its best results. In particular,
traffic volume is crucial for its efficiency. Daz \emph{et al.} in
\cite{mixmaster-reliable} prove that practical Mix designs achieve
results close to perfect indistinguishability for high traffic arrivals.
For each P2Priv transaction, the P2Priv protocol generates short but
numerous messages sent by random nodes. This can allow Mix-net to
assure high anonymity without unnecessary delays. The anonymity analysis
of Mix-net is independent of the current exposition and in our model
we assume perfect performance of Mix-net layer. We assume that the
adversary does not get any information about \emph{Alice} during the
establishment of $CC$. However, the adversary having the fraction
$\rho$ of malicious nodes can gain awareness that a particular content
is about to be distributed. If the cloning token describing this content
is passed via one of the malicious nodes, the adversary can disturb
$CC$ establishing in a way that allows him to get more information
in a subsequent phrase of content transportation. The adversary, who
looks for the initiator of particular content's distribution, can
try to narrow the circle of suspects. While possessing fraction $\rho$
of colluding nodes among $N$, he/she can break the cloning cascade
using the first malicious node which intercepts the cloning token.
The probability that the adversary manages to break $CC$ immediately
after sending out the token by \emph{Alice} equals $\Pr(\left|CC_{break}\right|=1)=\rho$
(we assume that colluding nodes are uniformly distributed among $N$).
The length of random-walk is one more hop longer with probability
$\Pr(\left|CC_{break}\right|=2)=(1-\rho)(p_{f}\,\rho+(1-p_{f}))$.
Then

\begin{eqnarray}
\Pr(\left|CC_{break}\right|=n) & = & (1-\rho)^{n-1}(p_{f}^{n-1}\,\rho+(1-p_{f})p_{f}^{n-2}).\label{eq:-1}\end{eqnarray}

As a result of this action the adversary concludes that the set of
parallel connections associated with the interesting content exchange
will contain an average of $\left|CC_{break}\right|$ links, where

\begin{eqnarray}
\left|CC_{break}\right| & = & \rho+\sum_{i=2}^{\left|CC\right|}i(1-\rho)^{i-1}(p_{f}^{i-1}\,\rho+(1-p_{f})p_{f}^{i-2})=\nonumber \\
 &  & \left[\right.(1+\left|CC\right|)p_{f}{}^{\left|CC\right|}(\rho-1)^{\left|CC\right|}-\nonumber \\
 &  & \left|CC\right|\mathit{p}_{f}{}^{\left|CC\right|+1}(\rho-1)^{\left|CC\right|+1}-\nonumber \\
 &  & \mathit{p}_{\mathit{f}}(\rho-2)+\mathit{p}_{f}{}^{2}(\rho-1)\left.\right]\nonumber \\
 &  & \mathit{p}_{\mathit{f}}(1+\mathit{p}\mathit{_{f}}(\rho-1))^{-1},\mbox{\ensuremath{\left|CC_{break}\right|\leq\left|N\right|}(1-\ensuremath{\rho})}.\label{eq:-2}\end{eqnarray}

After establishing of $CC$, P2Priv protocol starts direct, plain-text
connections from $CC$ members to destination nodes with shared content,
each independently delayed with a random interval of time. In the
analyzed P2P Scenario the content is shared between equivalent peers
\emph{$N$}. Then, having in mind the considered threat model, we
assume that malicious nodes are also uniformly distributed among destination
nodes. The adversary managed to reduce the cloning cascade from mean
length equal $\left|CC\right|$ to $\left|CC_{break}\right|$ (\ref{eq:-2}).
Next, he/she will eavesdrop on each content connection established
to $\rho\,\left|N\right|$ colluded nodes in order to detect transmission
of the content of his/her interest and in consequence in search of
connected $CC_{break}$ members. It immediately follows that the adversary
is able to identify $CC_{eavesdrop}$ peers connected towards the
particular content transportation, an average of 

\begin{center}
\begin{equation}
\left|CC_{eavesdrop}\right|=\rho\left|CC_{break}\right|.\label{eq:-7}\end{equation}

\par\end{center}

Until he/she can be certain that $CC_{eavesdrop}$ includes all of
the $CC$ or $CC_{break}$ members, he/she cannot be determine that
$CC_{eavesdrop}$ set includes \emph{Alice}. Let us estimate the uncertainty
of the adversary in linking of the found traces to the real initiator.
Each $CC_{eavesdrop_{k}}$ , $k=\{1,\ldots,\left|CC_{eavesdrop}\right|\}$
node can be \emph{Alice,} with equal probability 

\begin{equation}
p_{a1}=\Pr(CC_{eavesdrop_{k}}=Alice)=\left|CC_{break}\right|^{-1}.\label{eq:-4}\end{equation}

\emph{Alice }also can be outside eavesdropped nodes\emph{ }and\emph{
}can be each other honest node of the network in the number of $\left|N\right|-\rho\,\left|N\right|-\left|CC_{eavesdrop}\right|$.
The attack conducted by the adversary allows him/her to assign probability
that one of this nodes is \emph{Alice}, to each equal

\begin{equation}
p_{a2}=\frac{1-p_{a1}}{\left|N\right|-\rho\,\left|N\right|-\left|CC_{eavesdrop}\right|}.\label{eq:-5}\end{equation}

Then, we can stress the entropy of P2Priv in P2P Scenario as the following
sum of two components corresponding to the set of nodes managed to
have been eavesdropped by the adversary and the rest of nodes, respectively 

\begin{eqnarray}
\mathcal{H}_{\unitfrac{p2priv}{p2p}} & = & -\sum_{k=1}^{\left|CC_{eavesdrop}\right|}p_{a1}\log_{2}(p_{a1})\nonumber \\
 &  & -\sum_{l=1}^{\left|N\right|-\rho\,\left|N\right|-\left|CC_{eavesdrop}\right|}p_{a2}\log_{2}(p_{a2}).\label{eq:-9}\end{eqnarray}

Finally, after substitution and simplification we will then get 

\begin{eqnarray}
\mathcal{H}_{\unitfrac{p2priv}{p2p}} & = & \frac{\left|CC_{eavesdrop}\right|}{\left|CC_{break}\right|}\log_{2}(\left|CC_{break}\right|)-\nonumber \\
 &  & \frac{1-\left|CC_{eavesdrop}\right|}{\left|CC_{break}\right|}\log_{2}(\frac{1-\left|CC_{break}\right|}{\left|N\right|-\rho\,\left|N\right|-\left|CC_{eavesdrop}\right|}).\label{eq:-6}\end{eqnarray}

\begin{figure}
\begin{centering}
\includegraphics[scale=0.8]{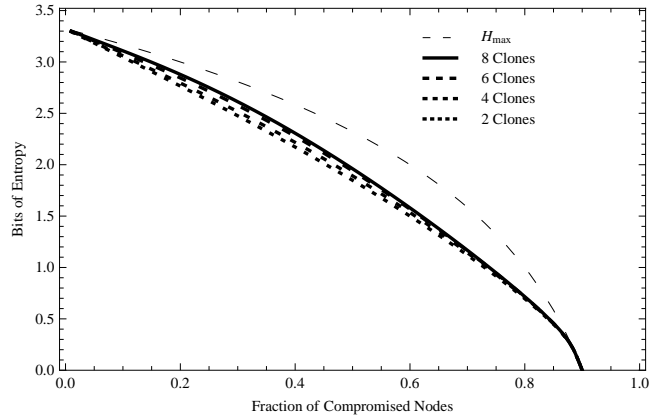}
\par\end{centering}

\caption{\label{fig:Entropy-of-P2P-P2PRIV-10}Entropy for P2Priv in P2P Scenario
as compared with maximum entropy, $\left|N\right|=10$.}

\end{figure}

Figure \ref{fig:Entropy-of-P2P-P2PRIV-10} shows entropy of P2Priv
architecture as applied to distributed P2P file-sharing service. The
entropy of small network ($\left|N\right|=10$) is plotted in the
full spectrum of colluding nodes. The presented results were obtained
for P2Priv in the following configurations $p_{f}=$ $\{1/2,2/3,4/5,6/7\}$,
which corresponds to mean cloning cascade lengths equal: 2, 4, 6,
and 8, respectively. We can observe that, even with a low number of
users, P2Priv achieves results close to the maximum in this distributed
service scenario and it is robust against a high fraction of compromised
nodes. 

\begin{figure}
\begin{centering}
\includegraphics[scale=0.8]{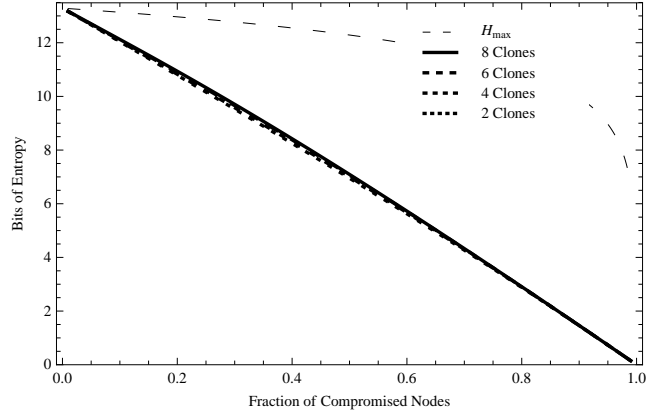}
\par\end{centering}

\caption{\label{fig:Enropy-of-P2P-P2PRIV-10 000}Entropy for P2Priv in P2P
Scenario as compared with maximum entropy, $\left|N\right|=10^{3}$.}

\end{figure}

Figure \ref{fig:Enropy-of-P2P-P2PRIV-10 000} investigates the robustness
of P2Priv in a large scale network comprised by $\left|N\right|=10^{3}$
nodes. We observe high entropy for a low-to-medium fraction of compromised
nodes. 

\subsection{\label{sub:P2PRIV-in-Client-Server}P2Priv in Client-Server Scenario}

Let us apply the P2Priv architecture to the opposite, centralized
service scenario. In this case, all user nodes $N$ connect to a single
server to receive a particular content, which is an \emph{item of
interest} of the adversary. We take into account that this server
is an easy target for observation and then assume that all its up-link
and down-link traffic is being eavesdroped on by the adversary.

\paragraph*{Single Request to Server.}

Similarly to previously analyzed attack scenario, we assume an active
adversary who is able to break cloning cascade $CC$ using fraction
$\rho$ of malicious nodes scattered in $N$. Summing it up, we can
stress that in this case, after an initiation of a content connection
to the server by \emph{Alice}, all associated connections will be
established with the server and eavesdropped on by the adversary.
Then, the number of all suspected nodes can be limited by the adversary
to $\left|CC_{break}\right|$ (\ref{eq:-2}). The adversary knows
that one of these nodes belongs to \emph{Alice}, each of them with
the probablity $p_{a1}=\left|CC_{break}\right|^{-1}$ (\ref{eq:-4}).
Then, entropy of P2Priv in Client-Server Scenarios is described as
follows

\begin{eqnarray}
\mathcal{H}_{\unitfrac{p2priv}{cs}} & = & -\sum_{k=1}^{\left|CC_{break}\right|}p_{a1}\log_{2}(p_{a1})=\nonumber \\
 &  & \log_{2}(\left|CC_{break}\right|).\label{eq:-3}\end{eqnarray}

\begin{figure}
\begin{centering}
\includegraphics[scale=0.8]{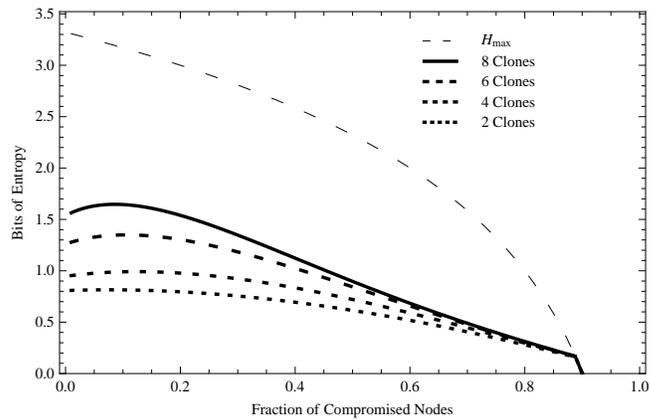}
\par\end{centering}

\caption{\label{fig:Entropy-of-CS-P2PRIV-10}Entropy for P2Priv in Client-Server
Scenario as compared with maximum entropy, $\left|N\right|=10$.}

\end{figure}

\begin{figure}
\begin{centering}
\includegraphics[scale=0.8]{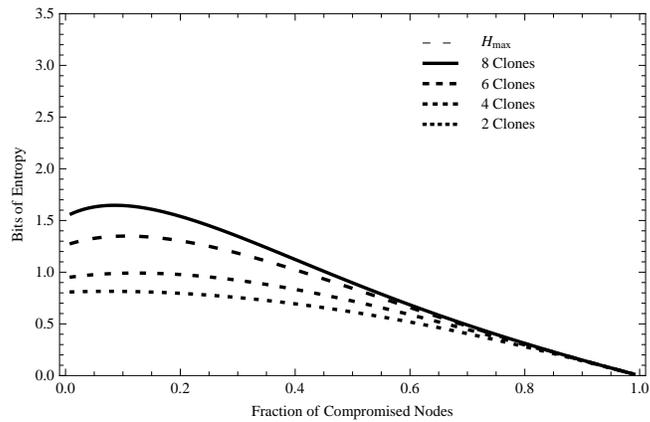}
\par\end{centering}

\caption{\label{fig:Entropy-of-CS-P2PRIV-10 000}Entropy for P2Priv in Client-Server
Scenario, $\left|N\right|=10^{3}$.}

\end{figure}

Figure \ref{fig:Entropy-of-CS-P2PRIV-10} and Figure \ref{fig:Entropy-of-CS-P2PRIV-10 000}
show entropy of P2Priv architecture applied for Client-Server services.
To allow an easier comparison between results obtained for small and
large scale networks, the range of entropy plotted in Figure \ref{fig:Entropy-of-CS-P2PRIV-10 000}
is the same as in Figure \ref{fig:Entropy-of-CS-P2PRIV-10}. Then,
it does not include plot for maximum entropy $\mathcal{H}_{\max}$
which certainly is the same as presented in Figure \ref{fig:Enropy-of-P2P-P2PRIV-10 000}.
As we expected, the results obtained in Client-Server Scenario are
much worse than those obtained in P2P environment for which P2Priv
was intended. Entropy of P2Priv architecture in centralized network
is independent of the network scale for a low-to-medium compromised
network and value about 1.5 bits for 10\%-20\% fractions of malicious
nodes.

\paragraph{Long-Term Observation. }

The previous analysis shows entropy of P2Priv in centralized service
scenario and the evaluated information leaks correspond to a single
connection to the server. However, considering client-server services,
sequential requests to a particular server are common. It should be
noticed, that P2Priv does not assure long-term availability of individual
nodes, so the $CC$ is deemed to change over time (with the exception
of the true sender). P2Priv was not designed for services characterized
by a sequential communication to a single network node and it cannot
be applied for these purposes in its current form.

\subsection{Summary of the Results}

We have found that the parallel transport architecture of P2Priv does
not assure a satisfactory anonymity level for centralized services.
Still the concept of transport parallelization and the moving of time-consuming
anonymization techniques to signalization layer seem to be very attractive
in the terms of anonymous traffic latency. In the rest of this paper
we will discuss the possibilities of using the parallelism-based approach
to network anonymization not limited to P2P content distribution.

\section{Network Privacy Preserving Parallel Topology}

Considering general purpose anonymous networks, we can distinguish
four basic types of network nodes : (i) client/user nodes, (ii) middle-man
nodes or proxy servers, (iii) the so called exit nodes, and (iv) service
nodes/servers. Client nodes send requests through middle-man nodes
to gain an anonymous access to services provided by service nodes
(iv). On the other hand, exit nodes (iii) are middle-man nodes which
are permitted by their policy to be boundary nodes of forwarding cascades---these
nodes connect directly to service nodes (iv) on behalf of users (i).
In various networks these sets are merged in different combinations.
In particular, all nodes of pure P2P networks can act as client nodes
(i), middle-man nodes (ii), exit nodes (iii), and server nodes (iv)
as necessary. Such division of roles corresponds to P2Priv. However,
when it comes to a generic traffic anonymization, we should separate
user nodes from exit nodes as not every user wishes to commit to sharing
his/her node as an exit node. Certainly, having in mind client-server
network applications, we should also distinguish service nodes (iv). 

To provide general-purpose sender anonymization we propose an anonymous
network architecture which joins P2Priv parallel transportation of
content (collective is comprised by parallel links, similary to P2Priv)
with proxy functions (each link is terminated by exit node). The new
architecture is derived from P2Priv P2P network and is dedicated to
general-puprose anonymous networks, though it will be referred to
as NetPriv (network privacy preserving parallel topology).

\begin{figure}
\begin{centering}
\includegraphics[scale=0.44]{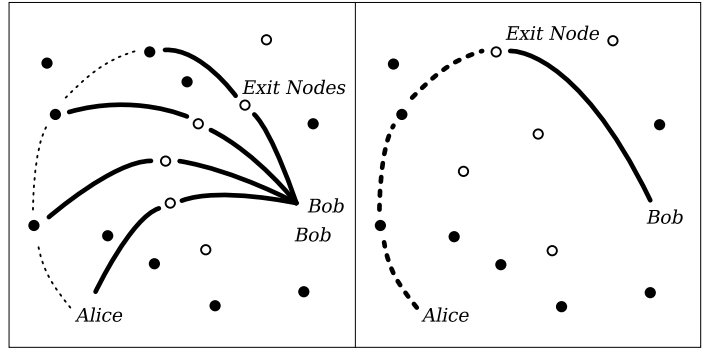}
\par\end{centering}

\caption{\label{fig:NetPriv-architecture-(left)}NetPriv architecture (\emph{left})
as compared to general-purpose classical anonymous network (\emph{right}). }

\end{figure}

Let us group nodes in the anonymous network as follows: $N$ is the
set of user nodes, potential initiators of communication and $E$
represents exit-nodes. We joined (i) and (ii) in our network model
as the anonymity of the proposed solution is basically based on the
difficulty of differentiation between real senders and other nodes
which simultaneously act the same way the senders act. NetPriv is
a hybrid solution which largely reflects P2Priv topology of parallel
links between $N$ nodes. However, each of these links is additionally
terminated by a link to a mixing exit node ($E$). Figure \ref{fig:NetPriv-architecture-(left)}
illustrates the model of the proposed network. In addition, it compares
it to classical anonymous networks.

To allow an anonymous communication in the described architecture
we propose the following sub-solutions: (i) persistent $CC$ path
selection by sender, (ii) time synchronization of requests sending
by $CC$ members.

\paragraph{Persistent $CC$ Path Selection.}

The discussion included in Section \ref{sub:P2PRIV-in-Client-Server}
shows that the parallel architecture of P2Priv, considered in the
scope of services characterized by series of user requests to the
same destination node or server, is not robust against long-term observations.
The reason is that path selection based on a random-walk does not
ensure persistent $CC$ paths, so the $CC$ is deemed to change over
time (with the exception of the true sender). The anonymity of $CC$
communications was from the beginning assured by the Mix-net layer.
Having this in mind, we propose a replacement of random-walk-based
path selection for $CC$ with free-route routing dedicated for Mix-network
\cite{disad-free-routes,sync-batching}. In this way a sender selects
one $CC$ for a whole session of an anonymous communication.

\paragraph{Requests Time Synchronization.}

The second issue the requires revision is answering the question how
exactly $CC$ members randomly delay their connections to destination
node/nodes. To assure that the adversary can not distinguish \emph{Alice}
from other $CC$ members by an analysis of connection times of individual
clones (clone that most frequently connects first can be distinguish
as the initiator), we propose inclusion of time field in cloning token
which specify starting time for connections and allows synchronization
of connection times. The time interval, indicated by \emph{Alice},\emph{
}should be longer than time required to send token via $CC$. Then,
each $CC$ member adds additinal randomly generated delay to time
indicated in received token and on time calcucated in such a way he
sends request to an exit node. Accordingly, cloning token will consist
of the following fields:

\[
\{dest\_addr,reqest\_time,request\}\]

and the request originated to a mixing exit node consists of:

\[
\{src\_addr,dest\_addr,request\}_{PK_{e}}.\]

An exit node sends request to a destination node or nodes, based on
information included in the received request. Certainly, a source address
in his request pointes at him.

\section{Anonymity Analysis for NetPriv}

Let us analyze anonymity of the new architecture. Previously, we have
shown that the parallel transport architecture assures high entropy
for distributed services of file-sharing (\ref{eq:-9}) without improvements
proposed in NetPriv. The vital question is how the new architecture
of NetPriv deals with Client-Server Service Scenarios (compare to
Section \ref{sub:P2PRIV-in-Client-Server}). As the exit nodes and
the user nodes have been disjointed, NetPriv anonymity model can allow
an analysis of a different fraction of compromised nodes in each of
these sets. One can imagine that different service scenarios provoke
and permit for different collaboration possibilities, especially as
it comes to the set of exit nodes which can be, e.g., dedicated servers
of a particular anonymous service, public proxies, or nodes voluntarily
provided by network users. As previously assumed $\rho$ represents
fraction of macilious user nodes. Let $\rho_{e}$ represent fraction
of malicious nodes among exit nodes. Then

\begin{center}
\begin{equation}
\left|CC_{eavesdrop}\right|=\rho_{e}\left|CC_{break}\right|.\label{eq:-10}\end{equation}

\par\end{center}

Similarly to the previous model, the adversary can assign probability
$p_{a1}$to each node of this set (\ref{eq:-4}). \emph{Alice} can
aslo be outside of eavesdropped nodes. We assume that destination
nodes/servers are compromised and $\rho_{e}$ of exit nodes is compromised.
Then each honest node of $N$ nodes can be \emph{Alice} with probability 

\begin{equation}
p_{a3}=(1-\frac{\left|CC_{eavesdrop}\right|}{\left|CC_{break}\right|})\frac{1}{N-\rho N-\left|CC_{eavesdrop}\right|}.\label{eq:-11}\end{equation}

Finally, the anonymity of the NetPriv architecture in the Client-Server
scenario is described by entropy

\begin{eqnarray}
\mathcal{H}_{NetPriv} & = & -\left|CC_{eavesdrop}\right|p_{a1}\log_{2}(p_{a1})-\nonumber \\
 &  & (N-\rho N-\left|CC_{eavesdrop}\right|)p_{a3}\log_{2}(p_{a3}),\label{eq:-12}\end{eqnarray}

\begin{eqnarray}
\mathcal{H}_{NetPriv} & = & \frac{\left|CC_{eavesdrop}\right|}{\left|CC_{break}\right|}\log_{2}(\left|CC_{break}\right|)-\nonumber \\
 &  & (1-\frac{\left|CC_{eavesdrop}\right|}{\left|CC_{break}\right|})\log_{2}(\frac{\left|CC_{break}\right|-\left|CC_{eavesdrop}\right|}{\left|CC_{break}\right|(N-\rho N-\left|CC_{eavesdrop}\right|)}).\label{eq:-13}\end{eqnarray}

\begin{figure}
\begin{centering}
\subfloat[$\left|N\right|=10$]{\includegraphics[scale=0.52]{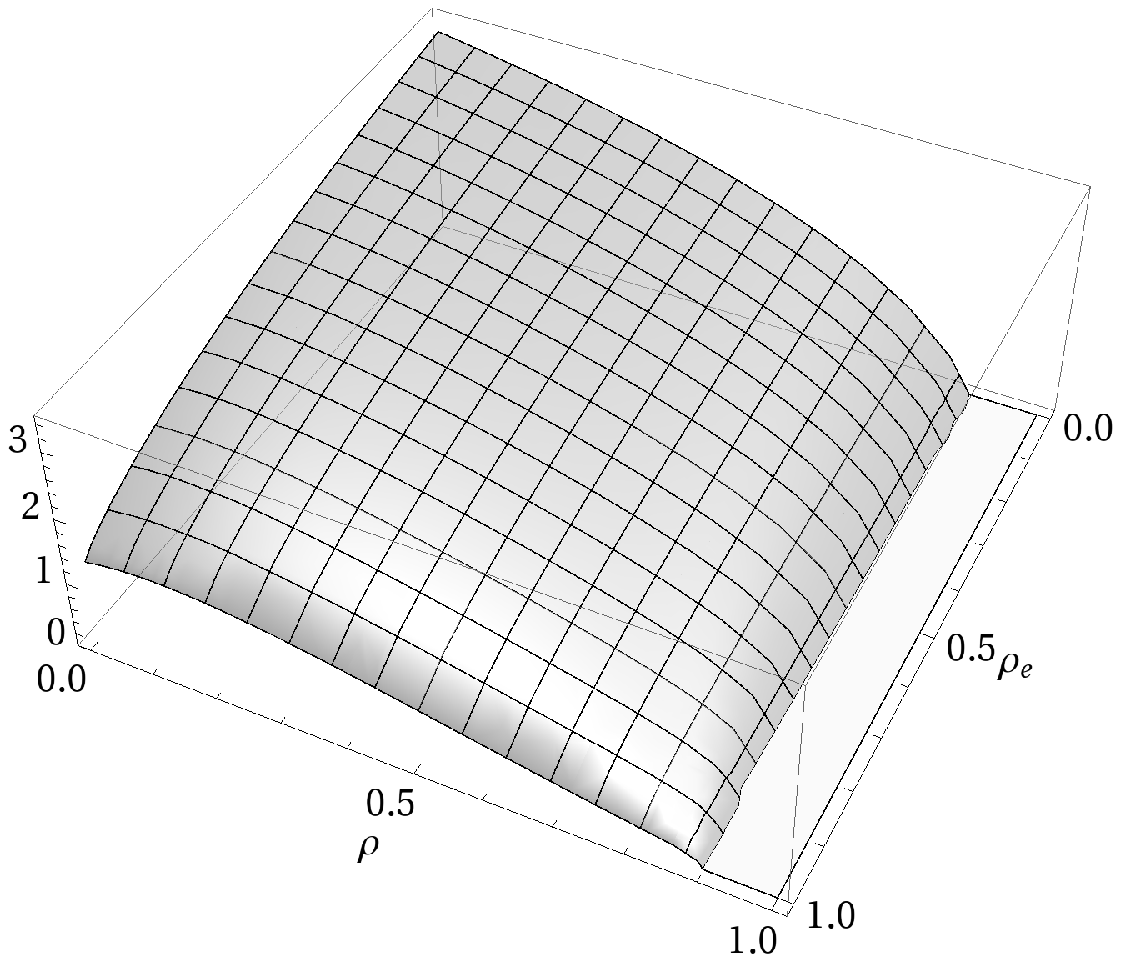}}\subfloat[$\left|N\right|=10^{3}$]{\includegraphics[scale=0.52]{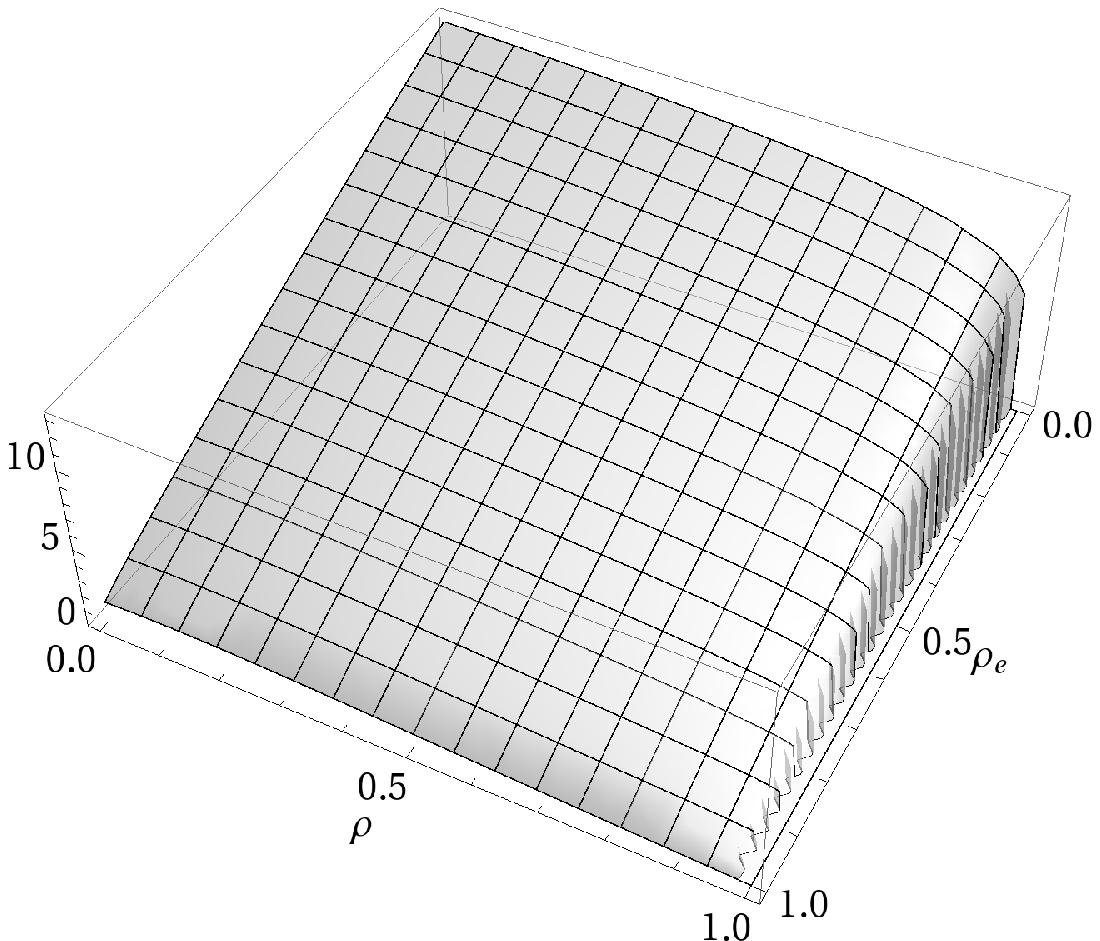}}
\par\end{centering}

\caption{\label{fig:Entropy-for-NetPriv}Entropy for NetPriv architecture,
$\left|CC\right|=4$.}

\end{figure}

Figure \ref{fig:Entropy-for-NetPriv} shows entropy of NetPriv in
the full spectrum of compromised user nodes ($\rho$) and compromised
exit nodes ($\rho_{e}$) for client-server services.

\begin{figure}
\begin{centering}
\includegraphics[scale=0.8]{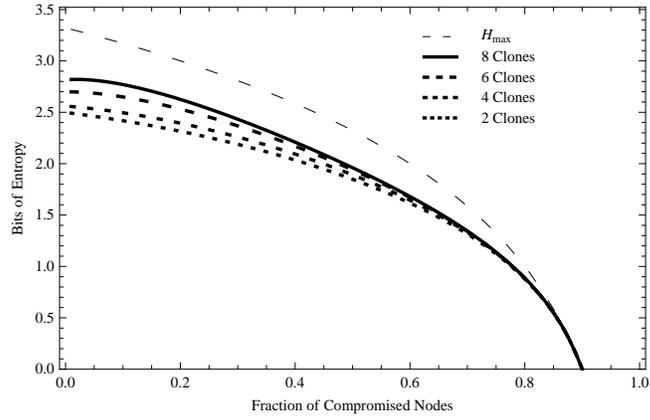}
\par\end{centering}

\caption{\label{fig:Entropy-of-NetPriv}Entropy for NetPriv in Client-Server
Scenario as compared with maximum entropy, $\left|N\right|=10$.}

\end{figure}

\begin{figure}
\begin{centering}
\includegraphics[scale=0.8]{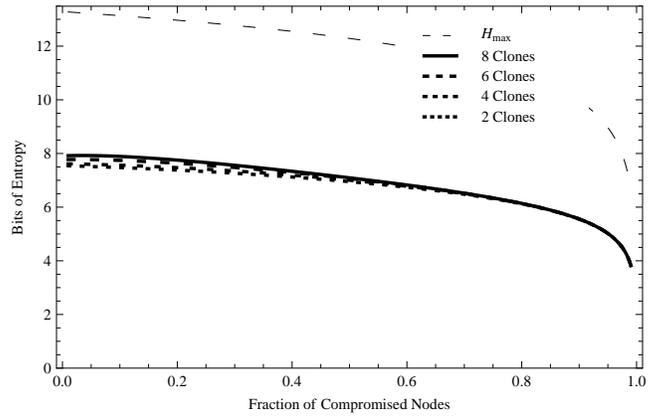}
\par\end{centering}

\caption{\label{fig:Entropy-of-NetPriv-1}Entropy for NetPriv in Client-Server
Scenario as compared with maximum entropy, $\left|N\right|=10^{3}$.}

\end{figure}

We can observe that the new architecture assures high entropy both
for small as well as large scale networks. Exit nodes can be particularly
vulnerable to being compromised. Figures \ref{fig:Entropy-of-NetPriv},
\ref{fig:Entropy-of-NetPriv-1} show entropy for NetPriv for constant
value of $\rho_{e}=1/2$. The results show that even for this high
fraction of collaborating exit nodes NetPriv assure high level of
anonymity.

\section{Conclusions and Future Work}

Today's topologies of anonymous networks shape anonymous communications
into hard to trace paths hiding their origin or destination. Transportation
of anonymous content via pervasive paths composed of several hops
is in favour of anonymity. Still, it imposes a significant traffic
bottleneck. The phenomenon of the Internet virtualization and practical
possibilities that allow us today to deploy heterogeneous overlay
networks of the world-wide range encourage consideration of new network
topologies adapted to specific network services. In this paper we
have proposed a framework of a novel architecture for anonymous networks---the
network privacy preserving parallel topology (NetPriv). NetPriv hides
initiators of communications by parallelization of anonymous links.
The new approach is based on the premise of the anonymous P2P network
called P2Priv \cite{MarPi08}. Contrary to P2Priv, the new architecture
can be suited to the anonymization of general-purpose network communications.
The new solution moves time-consuming anonymization techniques into
a signalization layer and combines the primary transport parallelization
principle of P2Priv with proxy functions. Additionally, a persistent
selection of signalization paths and requests time synchronization
mechanisms have been proposed. We applied an information theoretic
entropy measurement model to evaluate anonymity of both architectures.
We analyzed anonymity of the previous (P2Priv) and the new (NetPriv)
architectures with a particular emphasis on centralized, client-server
service scenarios. We have found that P2Priv can assure high degree
of anonymity in the limited scope of network services. P2Priv is dedicated
to a large size content distribution and requires a distributed service
overlay network to assure high degree of anonymity. Additionally,
we have found that anonymity of P2Priv decreases when destination
points of communication are not distributed. Contrary to the previous solution, we have
found that the new architecture can be applied to the anonymization
of various network communications, including client-server services (e.g., anonymous Web access).
For a realistic scope of compromised network nodes, NetPriv anonymity
is close to maximum. From the user's point of view, the new solution
offers a high level of anonymity within only a single proxy node.

The new parallelism-based approach presented in this paper gives the
framework for parallel anonymous topologies. However, it should be
noticed that this discussion encourages further work. The first vital issue, that needs to be
addressed, is the design of follow-up extensions allowing
for a bi-directional anonymous communication which includes a receiver anonymity.
Secondly, an interesting area of NetPriv analysis is its traffic performance
evaluation. Certainly, NetPriv, which allows anonymous transportation
via a single proxy, can be considered as the solution able to significantly
improve the speed of an anonymous communication (the traffic analysis
of P2Priv demonstrates substantial decreases of the content transportation
time as compared with traditional networks \cite{MarPi08,P2PRIVtrans,IM-thesis}).
However, a detailed and practical analysis can be helpful in determination
of proper NetPriv configurations and its exact impact on network traffic.

\bibliographystyle{plain}
\bibliography{bibfile}

\end{document}